\documentstyle[aps,pra,epsfig]{revtex}
\newcommand{\HeT}{{\mbox{He(2$^3$S)} }}
\newcommand{\HeST}{{\mbox{He$\uparrow$(2$^3$S)} }}
\newcommand{\pcore}{\mbox{($n$p)$^5$}}
\newcommand{\config}{\mbox{\{($n$p)$^5$($n$+1)s\}}}
\newcommand{\configNe}{\mbox{\{($2$p)$^5$3s\}}}
\newcommand{\configNa}{\mbox{\{($2$p)$^6$3s\}}}
\newcommand{\meta}{{\mbox{metastable} }}
\newcommand{\ajm}[1]{\mbox{$\mid \alpha_{#1} j^{at}_{#1} m_{#1} \rangle$}}
\newcommand{\Netpt}{{\mbox{Ne($^3$P$_2$)}}}

\newcommand{\cfive}{{\mbox{$C_5$} }}
\newcommand{\csix}{{\mbox{$C_6$}}}
\newcommand{\deltacsix}{{\mbox{$\Delta C_6$}}}
\newcommand{\deltacfive}{{\mbox{$\Delta C_5$}}}
\newcommand{\deltacfivenom}{{\mbox{$\Delta C_5^{nom}$}}}
\newcommand{\NaT}{\mbox{Na$_2(^3\Sigma_u^+$)}}
\newcommand{\NaS}{\mbox{Na$_2(^1\Sigma_g^+$)}}
\newcommand{\half}{\mbox{$\frac{1}{2}$}}
\newcommand{\ketlv}{\mbox{$\mid (l^v=0)~(m_l^v=0)\rangle$}} 
\newcommand{\ketlc}{\mbox{$\mid (l^c=1)~m_l^c\rangle$}} 
\newcommand{\kets}[2]{\mbox{$\mid (s^{#1}=\half)~m_s^{#1}{#2} \rangle$}}
\newcommand{\klv}{\mbox{$\mid 0^v \rangle$}} 
 
\newcommand{\ks}[2]{\mbox{$\mid m_s^{#1}{#2} \rangle$}}
\newcommand{\GofR}{{\mbox{$\Gamma(R)$}}}
\newcommand{\Kelvin}{\mbox{K}}
\newcommand{\cmps}{\mbox{cm$^3$/s}}
\newcommand{\rs}{\mbox{$\eta$} }
\newcommand{\dval}{\mbox{$d_{v}$}}
\newcommand{\dcore}{\mbox{$d_{c}$}}

\begin{document}

\title{Limit on suppression of ionization in metastable neon traps
due to long-range anisotropy.}
\author{M.R.\ Doery \protect\footnote{Current address:
Physics Laboratory, National Institute of Standards and Technology,
Gaithersburg MD 20877, U.S.A.}, E.J.D.\ Vredenbregt,
S.S.\ Op de Beek,
H.C.W.\ Beijerinck and B.J.\ Verhaar}
\address{Physics Department, Eindhoven University of Technology,
PO Box 513, 5600 MB Eindhoven, The Netherlands}

\maketitle

\begin{abstract}
This paper investigates the possibility of suppressing the ionization
rate in a magnetostatic trap of metastable neon atoms by spin-polarizing
the atoms. Suppression of the ionization is critical for the possibility
of reaching Bose-Einstein condensation with such atoms. We estimate the
relevant long-range interactions for the system, consisting of electric
quadrupole-quadrupole and dipole-induced dipole terms, and develop
short-range potentials based on the Na$_2$ singlet and triplet
potentials. The auto-ionization widths of the system are also
calculated. With these ingredients we calculate the ionization rate for
spin-polarized and for spin-isotropic samples, caused by anisotropy of
the long-range interactions. We find that spin-polarization may allow
for four orders of magnitude suppression of the ionization rate for Ne.
The results depend sensitively on a precise knowledge of the interaction
potentials, however, pointing out the need for experimental input. The
same model gives a suppression ratio close to unity for metastable xenon
in accordance with experimental results, due to a much increased
anisotropy in this case.
\end{abstract}

\section{Introduction}
The experimental observation of Bose-Einstein Condensation (BEC) in 1995
by three different groups \cite{bec1,bec2,bec3}
has fueled renewed interest in
this subject, with many more groups trying to achieve the conditions
under which the BEC-transition occurs. Stated simply, one has to reach
the point where the nearest-neighbor distance between the atoms is of
the same order of magnitude as their de Broglie-wavelength $\Lambda$. In
mathematical terms, the condition to fulfill is \cite{huang}
\begin{equation} \label{eqBEC}
n \Lambda^3 \geq 2.61,
\end{equation}
with $n$ the atomic number density. Experimentally, this can be achieved
by evaporative cooling of atoms caught in a trap,  with typical final
conditions given by a temperature of  $T_c \approx$ 1 $\mu$K and a density of
$n \approx 10^{14}$ atoms/cm$^3$ \cite{bec1} for Na, and $T_c \approx$
170 nK at $n \approx 3\times 10^{12}$ atoms/cm$^3$ \cite{bec2} for Rb.

Both the efficiency of the cooling process and the final density
achievable depend on the rates of inelastic and elastic collisions in
the ultra-cold atom clouds used \cite{druten}. With alkali atoms,
inelastic loss in binary collisions between ground-state atoms is due
mostly to hyperfine-changing collisions, with rates on the order of
$10^{-15} \cmps$ for Rb \cite{boesten,mies}. These rates are
generallly low enough to allow condition (\ref{eqBEC}) to be reached
\cite{Csfootnote}. In addition, all the alkali atoms have closed optical
transitions from the ground state that are accessible with commercially
available lasers, allowing for initial trapping and cooling in a
magneto-optical trap (MOT) \cite{straten&metcalf}. As a result, BEC has
now been achieved in Na, Rb and Li, while work is in progress on K and
Cs.

In comparison, the properties of rare gas atoms \cite{raregasprop} seem
less promising. As with atomic hydrogen, laser cooling and trapping
starting from the ground state is not practical due to the very short
wavelengths required (75~nm for neon). All rare gas atoms have a
metastable first excited state, however, that does allow for optical
manipulation. For He, this is the \{(1s 2s) 2$^3$S\} state, while for Ne
($n$=2) through Xe ($n$=5) the appropriate state can be written
\{($n$p)$^5$ ($n$+1)s $^3$P$_2$\}.
Condensates of such atoms might show interesting new
phenomena related to the available electronic energy, such as collective
decay or ionization. In addition, an ``atom laser'' \cite{atomlaser} based
on such condensates may find applications that are not covered by alkali
atoms: because of the large internal energy of metastable rare
gas atoms, one might make the comparison with the high-energy photons
of an XUV optical laser as opposed to those from a HeNe-laser.
On the downside, this internal energy of \meta rare gas atoms
is always enough to allow for
ionization in binary collisions \cite{penningreview}. Consequently,
ionization becomes a major loss process in metastable atom traps, with
experimental rates reported to be as large as $5 \times 10^{-11} \cmps$
for \HeT
\cite{mastwijk}. Such high rates make achieving condition (\ref{eqBEC})
impractical using the techniques currently employed for alkali atoms.

For He, Shlyapnikov and co-workers \cite{fedichev,fedichev1} have shown, however,
that the ionization rate
can be very strongly  suppressed by spin-polarizing the atoms. In the
process
\begin{equation} \label{eqIon}
\HeST  + \HeST  \rightarrow \mbox{He(1$^1$S)}
 + \mbox{He$^+$(1$^2$S)} + \mbox{e}^- ,
\end{equation}
the initial state then has total electron spin S~=~2, while in the final
state spin can only be S~=~0,1. Because the ionization process conserves
electron spin, ionization is prohibited in the S~=~2 (quintet) states
and can proceed only after a magnetic interaction between the electrons
causes a spin-flip during a binary encounter. Shlyapnikov and co-workers
\cite{fedichev,fedichev1} found that at ultra-cold temperatures the
relatively weak spin-dipole or magnetic spin-spin interaction then
limits the ionization rate in spin-polarized metastable He samples to
the order of $10^{-14}$ \cmps\ \cite{fedichev,fedichev1}.

For the heavier rare gases Ne through Xe, the spin-conservation rule
holds also. Therefore, suppression of ionization by spin-polarization
should be feasible as well. An added complication, however,
is the fact that the
unfilled \pcore\ core now has p-character, as opposed to the He case,
where it has s-character. This causes an orientation
dependence of the electrostatic interaction potentials that is
present even at large internuclear separation. During a collision, this
causes the \pcore\ cores to reorient, which, through the spin-orbit
interaction, induces electron spin-flips. Since this interaction has an
electrostatic origin rather than the magnetic one that causes spin-flips
in \meta He collisions, one may expect it to have a much larger
influence and possibly limit the ionization rate for spin-polarized
samples to much larger values than for He. It is the purpose of this
paper to estimate these effects for trapped clouds of atoms.

Out of the various \meta rare gases, we concentrate on \meta neon
because this element is being used in experiments presently going on in
our laboratory \cite{neonexperiment}. In addition, the properties
necessary for our calculations are most readily available for this
species \cite{neonprops}. While for most alkali atoms, the properties of
the atom-atom interactions are known quite accurately
\cite{alkaliprops}, the situation is very different for the \meta rare
gases, so that we have to make our own estimates.

\section{Metastable neon ionization rates.}
\label{secIonSup}

The \config-configuration of any of the heavy \meta rare gases contains
four energy levels, with the $^3$P$_2$-state always the lowest-lying
state. Table \ref{tableStates} lists the properties of these states in
neon. The four atomic fine-structure states give rise to a total of 144
molecular states. In what follows we will concentrate on the 25 states
that connect asymptotically to \Netpt + \Netpt, since the $^3$P$_2$
state is the only state that is both \meta and has non-zero angular
momentum, thus allowing for magnetic trapping and spin-polarization.

In the following sections we develop auto-ionization widths, short-range
and long-range real potentials for the equivalent process of
Eq.\ (\ref{eqIon}) for \Netpt. We then use them in a standard
coupled-channels calculation \cite{scatteringtheory} of the scattering
matrix $\cal{S}$. For each (positive) initial kinetic energy, we express
the asymptotic solution matrix of the calculation as a linear
combination of Bessel and Neumann functions for the radial part,
multiplied by channel functions $\Psi_k$ denoting electronic and angular
momentum states. For $\Psi_k$ we have used states with well-defined
total angular momentum quantum number $P$ with projection  $M$ onto a
laboratory axis, rotational quantum number $\ell$, electronic
(molecular) angular momentum quantum number $j$ and parity $\Pi$,
denoted as
\begin{equation}
\mid \Psi_k \rangle = \mid \Pi j \ell P M \rangle.
\end{equation}
Since $\Pi$, $P$ and $M$ are conserved in a collision, this has the
advantage that the coupled equations can be separated into blocks
characterized by these quantum numbers, which can then be solved
separately. The $\cal{S}$-matrix for such a block will be denoted
$S_{\Pi P M}$. The spin-polarized and non-polarized channels
can be easily identified because all the non-ionizing, quintet states
have $j = 4$ exclusively.

For any given parity, only even $\ell$ ($\Pi = 1$) or odd $\ell$ ($\Pi =
-1$) contribute. Because the \Netpt-atoms we are dealing with are
bosons, in addition only states that are symmetric under the exchange of
two atoms can contribute. This translates into the rule that $\ell + j$
must be even. Since parity is conserved in a collision, it follows that
the non-ionizing quintet states ($j = 4$) can only couple to other even
$j$ = 0,2 states.

Because of the presence of ionization, the $\cal{S}$-matrix for
(in)elastic scattering is not unitary.
Ionization rates follow from the loss of flux described by this
property. For the case of a trapped cloud of atoms, the distribution
of the direction of the initial relative velocity of two atoms is
spatially isotropic. Taking this into account, we find
for the ionization rate $K^{\text{pol}}$ in a polarized atom
sample
\begin{eqnarray}
K^{\text{pol}} & = & \sum_{\ell=0}^{\infty} K^{\text{pol}}_\ell\\
& = & \frac{2\pi}{9k^2} \sum_{\ell=0,even}^{\infty}
              \sum_{P=\mid j-l\mid}^{\mid j+l \mid} (2P+1)
              \{ 1- \sum_{j^\prime} \sum_{\ell^\prime}
              \mid S_{(\Pi=1) P M} (j^\prime \ell^\prime \leftarrow j \ell)
              \mid^2 \ \}
\end{eqnarray}
where the electronic angular momentum quantum number is limited
to $j = 4$.
Similarly, the rate for an unpolarized sample is given by
\begin{eqnarray}
K^{\text{unpol}} & = & \sum_{\ell=0}^{\infty} K^{\text{unpol}}_\ell\\
& = & \frac{\pi}{50k^2} \sum_{\ell=0}^{\infty}
              \sum_{j=0}^{4} [1 + (-1)^{\ell-j}]^2
              \sum_{P=\mid j-\ell\mid}^{\mid j+\ell \mid} (2P+1)
              \{1 - \sum_{j^\prime} \sum_{\ell^\prime}
              \mid S_{(\Pi=(-1)^\ell) P M}
              (j^\prime \ell^\prime \leftarrow j \ell) \mid^2\ \}.
\end{eqnarray}
A suppression ratio \rs may then be defined as
\begin{equation}
\rs = K^{\text{unpol}}/K^{\text{pol}}.
\end{equation}

\section{Long-range potentials}
\label{SecLongRange}
For two atoms with \config-configurations the lowest order terms in the
electrostatic interaction are the quadrupole-quadrupole (QQ) and
dipole-induced dipole (DD) terms \cite{c5ref,multipole}, with long-range
behavior
\begin{eqnarray} \label{eqElec}
V_{QQ} &=& C_5 R^{-5}\\
V_{DD} &=& C_6 R^{-6}.
\end{eqnarray}

The QQ-term is due entirely to the quadrupole moment of the tightly
bound \pcore-core and is therefore relatively small. Its order of magnitude
is given by
\begin{equation}
C_5 \approx \frac{q^2}{4\pi\varepsilon_0} \langle r^2_h \rangle^2.
\end{equation}
with $\langle r^2_h \rangle$ the
expectation value of the square of the orbital radius of the hole-state
in the \pcore-core
and $q$ the electron charge. Using electron wavefunctions calculated by
Hausamann \cite{Hausamann} we have found $\langle r^2_h \rangle \approx
1.0$~a.u., leading to $C_5 \approx 1$ a.u. Numerically, we have
implemented the complete state-dependence of $C_5$ based on the theory
of Refs. \cite{c5ref,multipole}, evaluated over a basis of product
states $ \ajm{1} \ajm{2} $, where $j^{at}$ is the atomic
electronic angular momentum
corresponding to the atomic state labeled by $\alpha$, and $m$ its
projection onto the internuclear axis. The maximum and minimum values of
$C_5$ that occur in the \Netpt+\Netpt\ subset are 0.47 a.u.\ and -0.37
a.u., respectively. These values will increase when going to the heavier
rare gases Ar, Kr and Xe, in accordance with the increasing value
of the $\langle r^2_h \rangle$-matrix element, which we have
listed in Table \ref{tabledD} (taken from Ref.~\cite{froese}).
According to the Table, for Xe the QQ-term will be about $26$
times larger than for Ne.

To obtain an estimate of the DD-term, we have used a second-order
perturbation treatment of the DD-interaction based on experimental
values \cite{neonprops} for the \config ~to $\{(n$p$)^5 (n+1)$p$\}$
transition,
analogous to our previous treatment of the DD-interaction for the
$\{(n$p$)^5 (n+1)$s$\}$+$\{(n$p$)^5 (n+1)$p$\}$ configuration in the
rare gases
\cite{ourpra}. Here we
have included all possible product states of all
$\{(n$p$)^5 (n+1)$s$\}$ and $\{(n$p$)^5 (n+1)$p$\}$ states,
which may limit the accuracy of the resulting \csix\
values to 30-40\% \cite{marinescuprivate}. The orientation-dependence
of the values obtained in
this way can be understood from a semi-classical description of the
DD-interaction, where the polarizability of a Ne atom is thought to be
made up of a combination of a large, spatially isotropic value $\dval
\approx 153$~a.u.\ due essentially to the $(n+1)s$ valence electron, with
an added small term $\dcore \approx 1.33$~a.u.\ determined by the
polarizability of the core. Only this smaller term depends on the
orientation of the \pcore-core, in the way described by Bussery and
Aubert-Frecon \cite{bussery}. Such a
model results in \csix\ values for the product state $ \ajm{1} \ajm{2} $
that (for $j^{at}_1 = j^{at}_2 = 2$) conform to
\begin{equation} \label{eqCsix}
\csix \propto \dval^2 + \frac{1}{18} \dcore \dval (34 - m_1^2 - m_2^2),
\end{equation}
to first order in \dcore, which is born out by our calculated
values. Since $\dcore \approx 0.01 \dval$
for Ne, the magnitude of the orientation-dependent term, 4.5 a.u., is only
0.25\% of the average value \csix\ = 1953 a.u. Again, this fraction increases
when going to heavier rare gases, finally attaining 1.3 \% for Xe (see
Tables \ref{tableMolec} and \ref{tabledD}). For He, Eq.\ (\ref{eqCsix}) is still appropriate
for the limiting value $\dcore = 0$.

Table \ref{tableMolec} lists the properties of the 25 molecular eigenstates
that diagonalize the combination of the DD- and QQ-interaction matrices with
the atomic energy-matrix. From the \cfive and \csix\ values listed, it
becomes clear that, on average, the QQ-term is quite a small effect. For
one it is overwhelmed by the DD-term out to internuclear distances of
$R \approx$ 4000 a.u. Beyond that point, its value is considerably less
than 1 pK, far below the collision energy currently obtainable in atom
traps. Therefore, the low-energy behavior of the the scattering rates
will be determined by the DD-term. Since, however, the anisotropy due to
the QQ-term (1 a.u.\ /$R^5$) is of the same order of magnitude as the
QQ-term itself, while that due to the DD-term (4.5 a.u.\ /$R^6$) is only
a fraction of its average value, the QQ term should still be expected to
determine the absolute value of the ionization rates for spin-polarized
atoms, which  are governed by inelastic, spin-changing processes.

\section{Short-range potentials} \label{secPots}

As mentioned in the introduction, there is little information available
about the interaction potentials of the heavier \meta rare gases. We can
obtain rough estimates of these, however, by considering the analogy
with ground-state alkali-atoms. For the case of Ne\configNe+Ne\configNe,
the appropriate choice is to base an estimate on the potentials of the
system Na\configNa+Na\configNa, consisting of the \NaT\ and \NaS\ states
that have been studied by a number of authors \cite{napots}.

Following Hennecart et al. \cite{hennecart1,hennecart}, we can expand each of
the four Ne\configNe\ atomic states (labeled by $\alpha = 1 \dots 4$) on a
basis of single-electron valence- (superscript $v$) and core-states
(superscript $c$),
\begin{eqnarray} \label{eqExpand}
\ajm{} & = &
\sum_{m_l^c=-1}^1 \sum_{m_s^c=-1/2}^{1/2} \sum_{m_s^v=-1/2}^{1/2}
c_{m_l^c,m_s^c,m_s^v}^\alpha \\
&& \times \ketlc \kets{c}{} \ketlv \kets{v}{}.
\end{eqnarray}
In the following we will write $\klv $ as short
notation for $\ketlv$ and leave out quantum numbers that do not
change. If we write the products of the valence electron states
in terms of triplet and singlet states,
\begin{eqnarray}
\mid T_{+1} \rangle &=& \klv_1 \klv_2 \ks{v}{=+\half}_1 \ks{v}{=+\half}_2 \\
\mid T_{-1} \rangle &=& \klv_1 \klv_2 \ks{v}{=-\half}_1 \ks{v}{=-\half}_2 \\
\mid T_0 \rangle &=& \frac{1}{\sqrt{2}} \klv_1 \klv_2 \times
          \left( \ks{v}{=+\half}_1 \ks{v}{=-\half}_2 + \right.
          \left. \ks{v}{=-\half}_1 \ks{v}{=+\half}_2 \right)\\
\mid S \rangle &=& \frac{1}{\sqrt{2}} \klv_1 \klv_2 \times
          \left( \ks{v}{=+\half}_1 \ks{v}{=-\half}_2 - \right.
          \left. \ks{v}{=-\half}_1 \ks{v}{=+\half}_2 \right),
\end{eqnarray}
then the Na$_2$ potentials correspond to matrix elements of the atom-atom
interaction $V_{\text{Na}}$ given by:
\begin{eqnarray}
\NaT & : & \langle T_i \mid V_{\text{Na}} \mid T_j \rangle \delta_{i,j} \hspace{2mm}
(i,j=-1,0,1)\\
\NaS & : & \langle S \mid V_{\text{Na}} \mid S \rangle.
\end{eqnarray}
From this we can construct Ne$_2$ potentials if we multiply
$V_{\text{Na}}$ by a unit core-factor that is diagonal in all core
quantum numbers.
Such potentials should contain the essential features of the
valence-electron exchange interactions, but do not describe the
interaction of the core hole of one atom with the valence electron of
the other. Since the radius of the valence electron's orbital is
considerably larger than that of the core-electrons \cite{ddvalues},
this approximation should be reasonable. Also, core-valence electron
terms can be added in a semi-empirical way. For an accurate description
of the wells and turning points, however, either {\em ab-initio}
potentials or experimental data would be required.

\begin{figure}
\epsfig{file=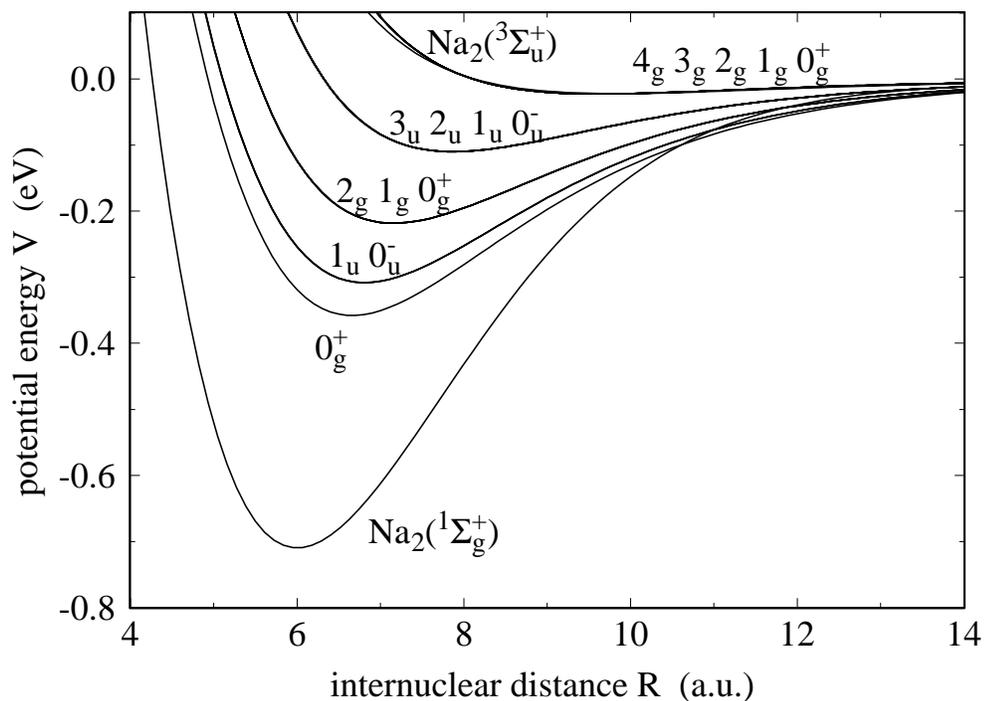,width=13cm}
\caption{\label{figPots} The adiabatic potentials of \NaT and
\NaS \protect\cite{napots},
and the short-range, adiabatic, \Netpt+\Netpt-potentials
derived from them with the formalism of Sec.\ \ref{secPots}.
The potentials of the Ne$_2$ quintet states (characterized
by $j=4$ and $S=2$) coincide with those of \NaT.
The molecular potentials are labeled by their Hund's case ($c$)
classification.
} 
\end{figure}

Figure (\ref{figPots}) displays diagonalized \Netpt+\Netpt\ potentials
developed in this way, together with the \NaS\ and \NaT\ potentials on
which they are based. The 9 possible, fully spin-polarized, Ne$_2$
states that are characterized both by a total angular momentum $j = 4$
and a total electron spin $S = 2$, exactly coincide with the \NaT\
states. The repulsion of the valence electrons leads to a turning point
$R_{\text{tp}} \approx 8$ a.u.\ for vanishing asymptotic kinetic energy
in this case.

All states conform to Hund's case ($c$) \cite{germanspec}, and in
addition can be characterized by their total angular momentum $j = 0
\dots 4$. Consequently, their $g/u$-symmetry follows immediately from $j$
as (-1)$^j$. The well depth of these Ne$_2$ states increases with
decreasing $j$, since more and more \NaS-character becomes mixed in. The
state with the deepest well, of symmetry $0_g^+$, does not coincide with
the \NaS\ state since it is not a pure singlet. In order for deeper
wells to occur, the spin-orbit coupling of each of the Ne atoms must be
allowed to compete with the \NaS-\NaT\ splitting, which requires mixing
with states correlating to other asymptotes than \Netpt+\Netpt\ from
within the same configuration \configNe+\configNe. We will discuss
possible consequences of including only the \Netpt+\Netpt-configuration
in section \ref{secConclusion}.

\section{Auto-ionization widths}

In scattering calculations, ionization can be introduced by adding
an imaginary part to the real-valued atomic interaction potentials
\cite{penningreview,exactref}, creating
a complex ``optical'' potential $W$,
\begin{equation}
W(R) = V(R) - i \GofR/2,
\end{equation}
with \GofR\ the so-called auto-ionization width.

Op de Beek et al.\ \cite{opdebeek,opdebeek2} have described a numerical
method for
the calculation of \GofR, based on a recurrence recipe developed by Rico
et al. \cite{rico}. Briefly, one calculates matrix elements between
initial and final states of the Coulomb interaction between the
individual electrons involved in the ionization process. In the process
considered here,
\begin{equation} \label{eqIonNe}
\Netpt  + \Netpt  \rightarrow \mbox{Ne($^1$S)}
 + \mbox{Ne$^+\{(2p)^5$ $^2$P)\}} + \mbox{e}^- + \mbox{11.7 eV},
\end{equation}
the dominant ionization mechanism is the so-called {\em exchange} mechanism
\cite{radiationnote}.
This means that the valence electron of atom 1 transfers to the available
core-state of atom 2, leaving the core of atom 1 to form a positive ion.
The valence electron of atom 2 is now no longer bound and leaves the
scene. In the
basis of valence-electron and core-states introduced in Eq.\ (\ref{eqExpand}),
a partial ionization amplitude $\gamma$ may be defined, given by
\begin{equation} \label{eqGofR}
\gamma = \int \mbox{d}\vec{r}_1 \mbox{d}\vec{r}_2
{\Psi^v_1} (\vec{r}_1) {\Psi^c_2}(\vec{r}_1)^*
\frac{q^2}{4\pi\epsilon_0 \mid \vec{r}_1 - \vec{r}_2 \mid}
\Psi^v_2 (\vec{r}_2) {\Psi^f}(\vec{r}_2)^*
\end{equation}
with $\vec{r}_1$ and $\vec{r}_2$ the position vectors of the active
electrons of atom 1 and 2, respectively.
The various single-electron valence and core states can be characterized
by their
orbital and spin quantum numbers with corresponding projections on the
internuclear axis,
\begin{eqnarray}
\Psi^v_1 &\propto& \mid {l^v_1} = 0; {(m_l)^v_1} = 0 \rangle
\mid {s^v_1} = \half; {(m_s)^v_1} \rangle \nonumber \\
\Psi^v_2 &\propto& \mid {l^v_2} = 0; {(m_l)^v_2} = 0 \rangle
\mid {s^v_2} = \half; {(m_s)^v_2} \rangle  \\
\Psi^c_2 &\propto& \mid {l^c_2} = 1; {(m_l)^c_2} \rangle
\mid {s^c_2} = \half; {(m_s)^c_2} \rangle, \nonumber
\end{eqnarray}
while the free electron's wavefunction $\Psi^f$
is given by a Coulomb wave, characterized
by orbital angular momentum quantum number $\lambda$ with projection $\mu$,
multiplied by a spin state $\mid s_f = \half; m^s_f \rangle$.
Since the Coulomb interaction does not affect spin, Eq. (\ref{eqGofR})
implies a delta-function requiring
${(m_s)^v_1} + {(m_s)^v_2} = {(m_s)^c_2}  + m^f_s$.
In addition, conservation of the azimuthal part of
the orbital angular momentum requires
${(m_l)^v_1} + {(m_l)^v_2} = {(m_l)^c_2} + \mu^f$
so that $\mid \mu \mid \leq 1$.

By using the inverse of the expansion in Eq.\ (\ref{eqExpand}) on
atom 1 as well as atom 2,
$\gamma$ may be transformed into a molecular basis. From the resulting
molecular ionization amplitudes $\gamma^{mol}$, matrix elements of the
ionization width \GofR\
can then be constructed by summing over the available final states
(represented symbolically by $F$),
\begin{equation}
\GofR_{i,j} = \frac{2\pi}{\sqrt{2 E_e \hbar^2 /m_e}}
\sum_{F} \gamma^{mol}_{i,F} \gamma^{mol}_{j,F}
\end{equation}
where $i,j$ denote individual molecular states, $m_e$ is the electron mass
and $E_e$ the energy carried away by the electron ($E_e$ = 11.7~eV for the
process of Eq.\ (\ref{eqIonNe})).

\begin{figure} 
\epsfig{file=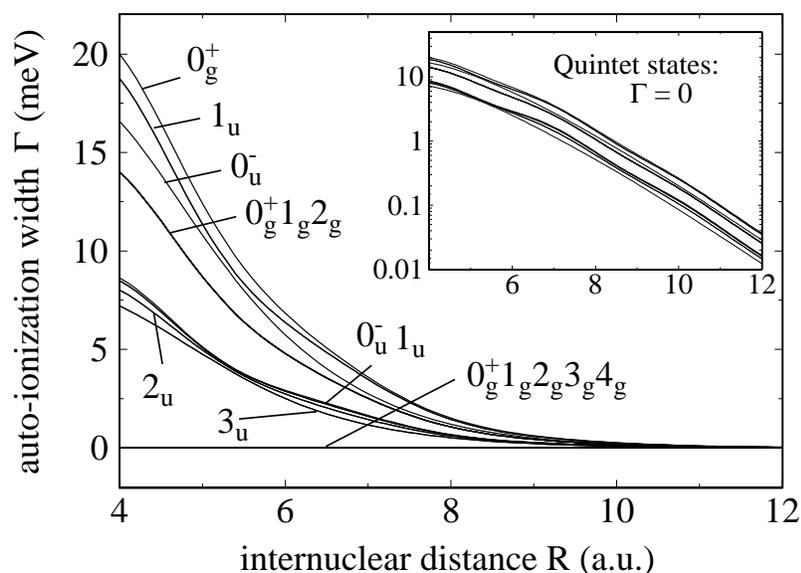,width=13cm}
\caption{\label{figWidths} 
Ab-initio auto-ionization widths $\Gamma$ 
for the \Netpt+\Netpt-system
as a function of internuclear distance $R$. The widths
are given in the same, adiabatic basis as used for 
Fig.\ (\ref{figPots}). The inset shows the widths on a
logarithmic scale. Note the vanishing widths of the quintet states.
} \end{figure}

We have calculated all elements of \GofR\ relevant for the ionization
process of Eq.\ (\ref{eqIonNe}), using electron wavefunctions tabulated
by Hausamann \cite{Hausamann}, and using a modified version of the
computer code developed by Op de Beek et al. \cite{opdebeek,opdebeek2}.
Figure (\ref{figWidths}) displays the results in the same basis as used
for Fig.\ (\ref{figPots}), for internuclear distances 4~$< R \mbox{(a.u.)}
<$~12. As expected, the quintet states all have zero widths. Also as one
would expect, the widths of the other states show an approximately
exponentially decreasing behavior with $R$. At each $R$, the ionization
width is largest for the states with the largest ($S = 0$)-component.

\section{Calculated ionization rates for metastable neon.}

\subsection{Threshold behavior.}

\begin{figure} 
\epsfig{file=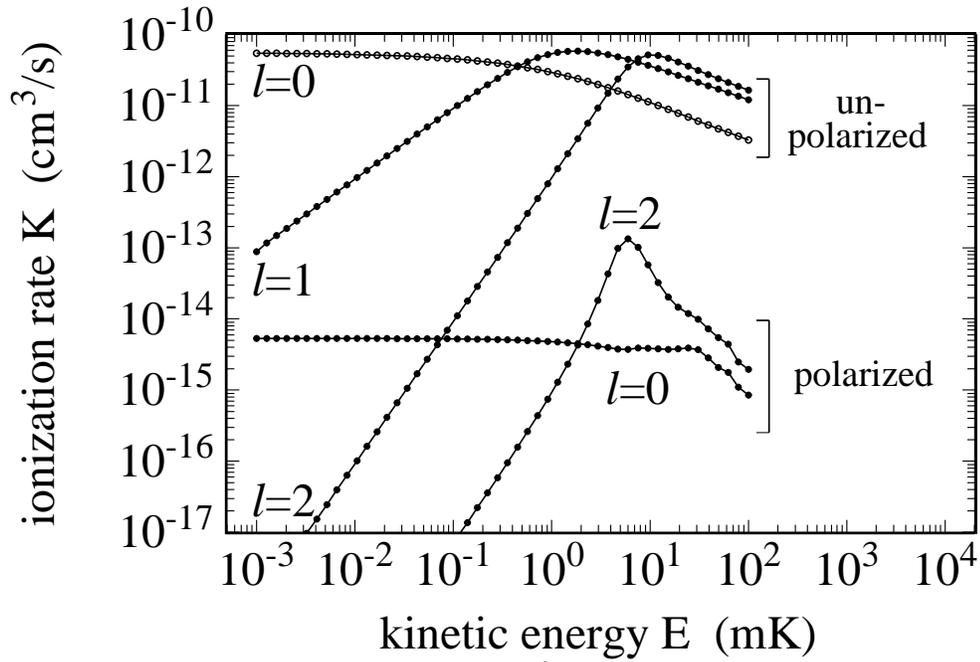,width=13cm}
\caption{\label{figKfull} 
Partial ionization rates 
for trapped \Netpt-atoms as a function of the
collision energy $E$. Rates are given for unpolarized
($K^{unpol}_{\ell}$) as well as for polarized atoms 
($K^{pol}_{\ell}$)
for angular momentum quantum numbers $\ell = 0 \dots 2$.
The rates show an obvious Wigner-law behavior. The rates
for polarized atoms are about $10^4 \times$ lower than those for
unpolarized atoms.
} 
\end{figure}

With the ingredients developed above, ionization rates $K_\ell$ for
polarized and unpolarized atom clouds were calculated as a function of
the collision energy $E$ at ultra-cold temperatures. Figure
(\ref{figKfull}) shows the result for $\ell = 0 \dots 2$ in the range $
1~\mu\Kelvin < E < 100~\mbox{m}\Kelvin$. The curves show the typical
behavior expected from Wigner's threshold law \cite{scatteringtheory}
for an exothermic process, $K_\ell \propto E^{\ell}$. At very low
temperature, the contribution for $\ell = 0$ always dominates, leading
to an essentially constant ionization rate in this regime. As discussed
in Sec. (\ref{secIonSup}), due to bose-symmetry
curves for odd partial waves are absent in the case of polarized atoms.

\begin{figure}
\epsfig{file=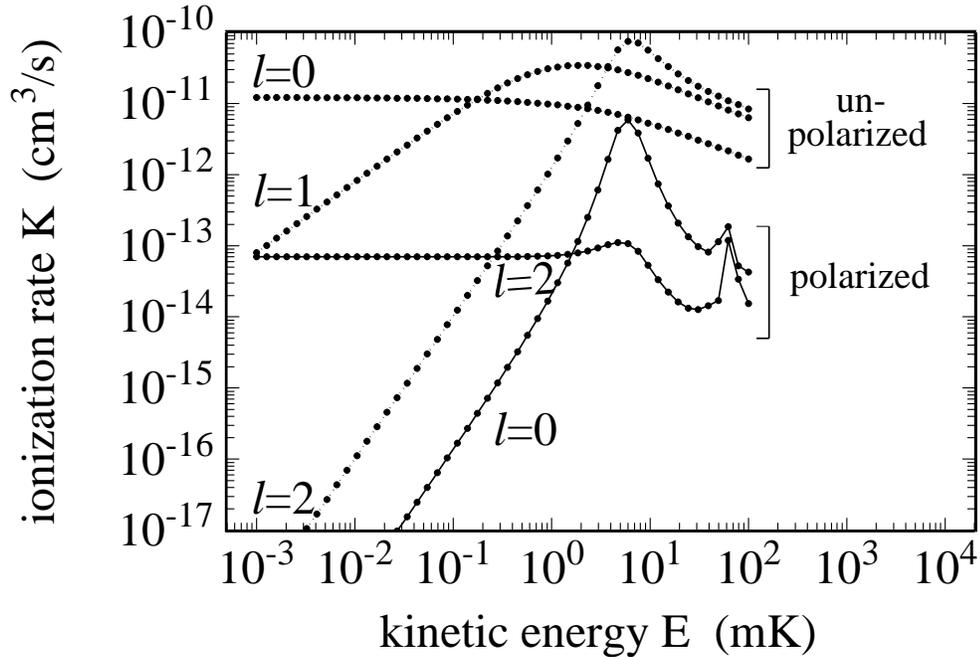,width=13cm}
\caption{\label{figKsingle} 
Partial ionization rates 
for trapped \Netpt-atoms as a function of the
collision energy $E$. In evaluating the potentials for the
system, the \NaS-potential was artificially replaced by the
\NaT\ potential, removing the large energy-splitting between
ionizing and non-ionizing states. The resulting
suppression ratio is much less than for the case
of Fig.\ (\ref{figKfull}).
} 
\end{figure}

The calculations of Fig.\ (\ref{figKfull}) show that a strong
suppression, $\rs \approx 10^{4}$, is still possible notwithstanding the
long-range anisotropy of the potentials. The resulting suppressed
ionization rate, $K^{\text{pol}} \approx 5 \times 10^{-15} \cmps$ at $E =
1~\mu\Kelvin$, would certainly allow one to reach the BEC transition,
being comparable to the depolarization rate for Rb, $K^{depol} = 1 \times
10^{-15} \cmps$ \cite{boesten,mies}. This result is largely due to the
fact that the depolarization due to the long-range anisotropy is
counteracted strongly at short range by the separation of the
non-ionizing quintet states from the other, ionizing states [see Fig.\
(\ref{figPots})]. This is illustrated in Fig.\ (\ref{figKsingle}), which
displays the rates that result from artificially replacing the
\NaS-potential by the one for \NaT\ in the calculation of the
\Netpt+\Netpt-potentials. In this case, no short-range separation
between ionizing and non-ionizing states occurs, with the result that
\rs is reduced to $\rs \approx 170 $. This shows that the short-range
behavior of the potentials is very important for the amount of
suppression achievable.

\begin{figure}
\epsfig{file=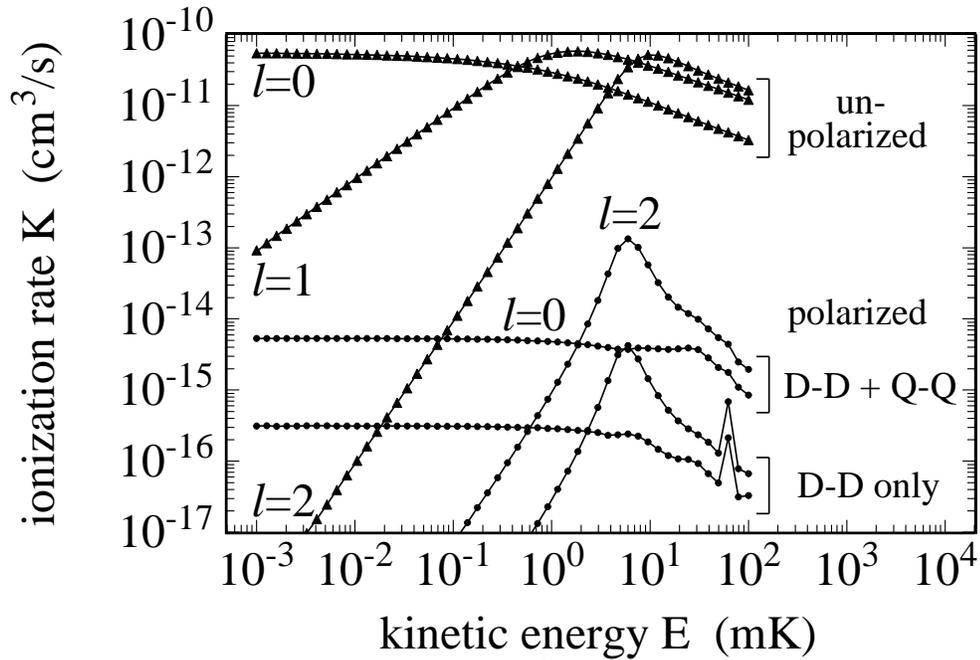,width=13cm}
\caption{\label{figKnoQQ}
Partial ionization rates for trapped \Netpt-atoms as a function of the
collision energy $E$, calculated including or excluding the effect of
the quadrupole-quadrupole (QQ) interaction on the potentials. For the
polarized case the results are labeled``DD + QQ'' and ``DD only'',
respectively. In general, these rates drop by about an order of
magnitude when the QQ-term is omitted. For the unpolarized case, the
solid lines give the result including, and the triangles the result
excluding the QQ-term. Little difference is observed in this case. }
\end{figure}

To investigate the relative importance of the quadrupole-quadrupole {\em
vs.\ }the dipole-dipole interaction, we have calculated the ionization
rates for polarized and non-polarized samples, with and without the
QQ-term in the long-range potentials. Fig.~(\ref{figKnoQQ}) shows the
results. From this figure, it becomes clear that the QQ-term is
the dominating influence on the ionization rate of spin-polarized
samples, which change by about an order of magnitude when the QQ-term is
removed. At the same time, however, removing the quadrupole term has
little influence on the threshold behavior of the rates nor on the
ionization rate for unpolarized samples, confirming the dominating role
of the dipole term on the elastic scattering already anticipated in
Sec.~\ref{SecLongRange}.

\subsection{Influence of quasi-bound states.}

\begin{figure} 
\epsfig{file=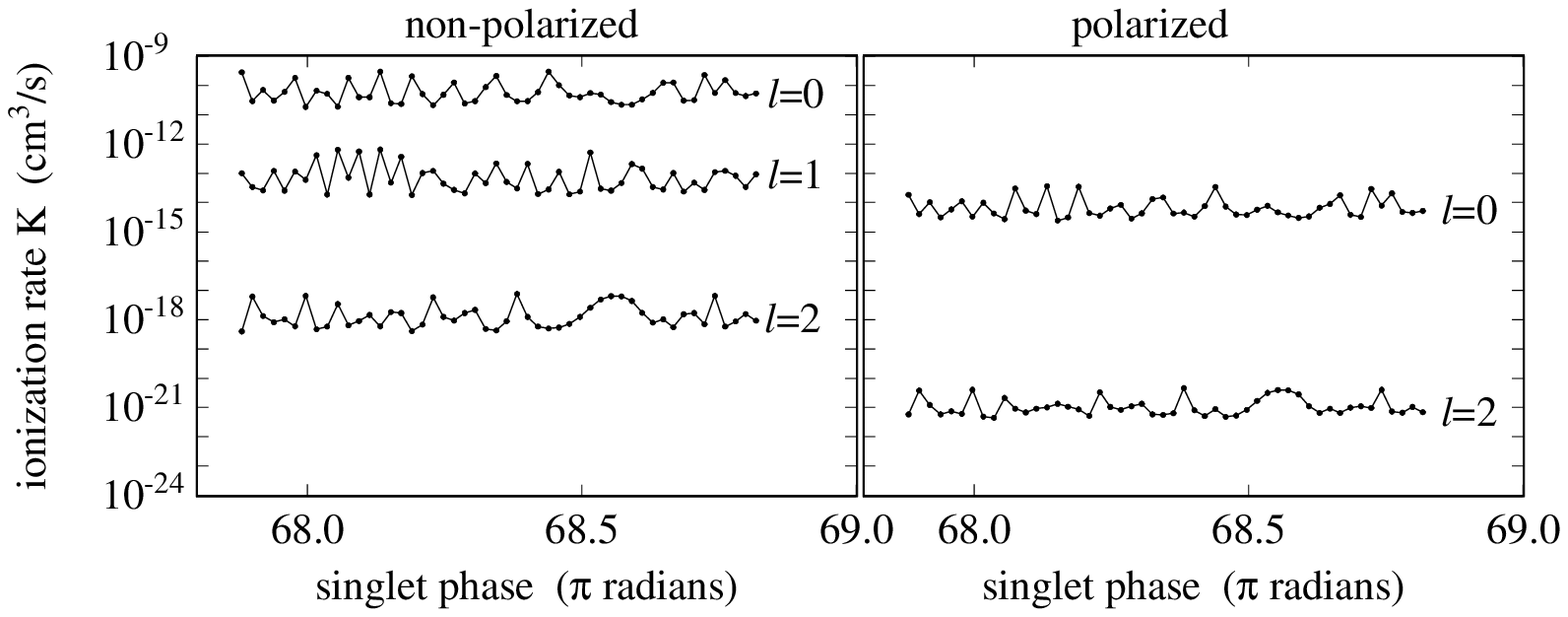,width=13cm}
\caption{\label{figVarS} 
Partial ionization rates 
for trapped \Netpt-atoms as a function of the
depth of the \NaS-potential from which the
potentials of the \Netpt+\Netpt-system are calculated.
The depth of the \NaS-potential is given in terms of
the semi-classical phase $\phi$ [see Eq.\ (\ref{eqPhase})].
Only relatively small variations in the rates are observed, 
indicating the minor influence of quasi-bound resonances in the
ionizing states on the ionization rate of the non-ionizing, polarized
quintet states. The collision energy is $E = 1 \mu\Kelvin$.
} 
\end{figure}

\begin{figure} 
\epsfig{file=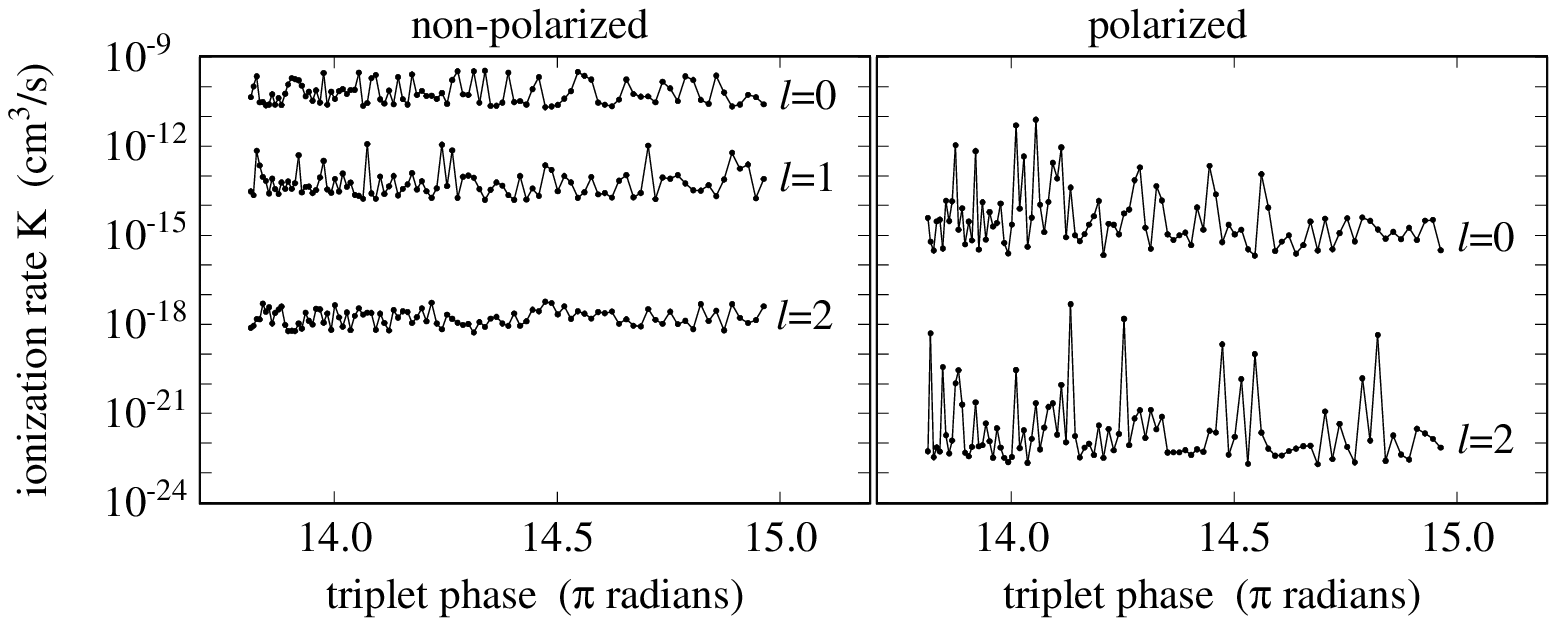,width=13cm}
\caption{\label{figVarT}
Partial ionization rates 
for trapped \Netpt-atoms as a function of the
depth of the \NaT-potential from which the
potentials of the \Netpt+\Netpt-system are calculated.
The depth of the \NaT-potential is given in terms of
the semi-classical phase $\phi$ (see Eq.\ (\ref{eqPhase})).
Very large variations in the rates are observed, 
indicating narrow quasi-bound resonances occurring in the
non-ionizing quintet states. These may adversely affect
the possiblities of achieving BEC with metastable rare gases.
The collision energy is $E = 1 \mu\Kelvin$.
} 
\end{figure}

Since the short-range potentials for the Ne-system are not well known,
we have to consider variations of the ionization rates due to the
uncertainty in the potentials. Of particular importance are resonances
due to quasi-bound states with total energy equal to that of the $\ell =
0$ quintet state. These can give rise to the formation of a relatively
long-lived collisional complex, in which the atoms undergo a number of
vibrations before they exit. In such a case, the turning point at small
$R$, where ionization is most likely, may be encountered several times,
leading to an enhanced ionization probability. To investigate the
importance of such resonances, we have added a Gaussian ``bump'' of
variable height to the bottom of the well of either the \NaT\ or the
\NaS\ potentials
from which the \Netpt+\Netpt-potentials are calculated. By increasing
the height of the bump, we decrease the total semi-classical phase $\phi$
available in the
well, effectively moving the ro-vibrational states it contains up in
energy. The phase $\phi$ is calculated from \cite{semiclassical}
\begin{equation} \label{eqPhase}
\phi = \int_{R_{\text{tp}}}^{\infty} \mbox{d}R \sqrt{-2\mu V(R)/\hbar^2}
\end{equation}
with $V(R)$ the potential under consideration and $R_{\text{tp}}$ its
classical turning point for vanishing asymptotic kinetic energy. Figures
(\ref{figVarS}) and (\ref{figVarT}) display the variation of
$K^{\text{pol}}_\ell$ and $K^{\text{\text{unpol}}}_\ell$ with the phase-content
of the \NaS\ and \NaT\ potentials obtained in this way. Since we vary
the phase-content by more than $\pi$, resonances are certainly present
in both cases.

Figure (\ref{figVarS}) shows that a variation of the \NaS\ singlet
potential only has a small effect on $K^{\text{pol}}$. In this case, only the
energy levels in the quintet states are unaffected by the change in
phase-content, since they derive solely from the \NaT\ triplet
potential. We can conclude, therefore, that resonances in states other
than the quintet states do not seriously alter $K^{\text{pol}}$. The most
likely explanation for this is that such resonances are extremely wide,
due to the large ionization probability in the non-quintet states, and
therefore correspondingly weak. In contrast, a variation of the \NaT\
potential depth has a very large effect on $K^{\text{pol}}$, as found from
Fig.\ (\ref{figVarS}). Here, atoms that initially have $\ell = 0$ in the
quintet, are coupled to an $\ell \neq 0$ quintet state at the same total
energy, which has a centrifugal barrier at $R \approx 75$~a.u. Such a
quasi-bound level may have a very sharp and therefore strong resonance,
since in this state, too, ionization is now prohibited. During its
strongly enhanced lifetime, the molecule may therefore make many
vibrations before it tunnels through the barrier or is transferred back
to the initial state. All the while, the anisotropy present in the
electrostatic interactions weakly couples the molecule to the ionizing
states, leading to a greatly enhanced $K^{\text{pol}}$. In our calculations,
$K^{\text{pol}}_{\ell=0}$ reaches a maximum value of $5 \times 10^{-11}
\cmps$ at resonance, at which point $\rs \approx 1$. Again, these values
would be prohibitive for reaching the BEC transition. Over the whole
phase-variation $\Delta\phi = \pi$, however, the suppression ratio
mostly takes values between $10^3$ and $10^4$, showing that a large
suppression is rather likely.

These results clearly indicate the extreme importance of an accurate
knowledge of the full potentials, or, equivalently, the scattering
lengths for the process under consideration. We expect that such
information will only be obtained reliably from an {\em ab-initio}
theoretical treatment of the short-range potentials in combination with
experimental spectroscopic determinations of the bound molecular states.
Until now, this information has only been available for some of the
alkali atoms. Since all metastable rare gas atoms other than He have
more than one bosonic isotope, one may hope to avoid such resonances by
an appropriate choice of isotope. Since there are a total of nine
quintet states, however, with as many as 5 different potential curves if
additional short-range interactions are considered, there may be so many
different resonances that avoiding them altogether is impossible.

\subsection{Variation with long-range anisotropy.}
\label{secVarC}

In order to make estimates of the suppression ratio for the heavier rare
gases Ar through Xe, we have to study the dependence of the ionization
rates on long-range anisotropy. According to Table~\ref{tabledD}, both
the anisotropy in the dipole-dipole term as well as that in the
quadrupole-quadrupole term becomes progressively larger going from Ne to
Xe.

\begin{figure}
\epsfig{file=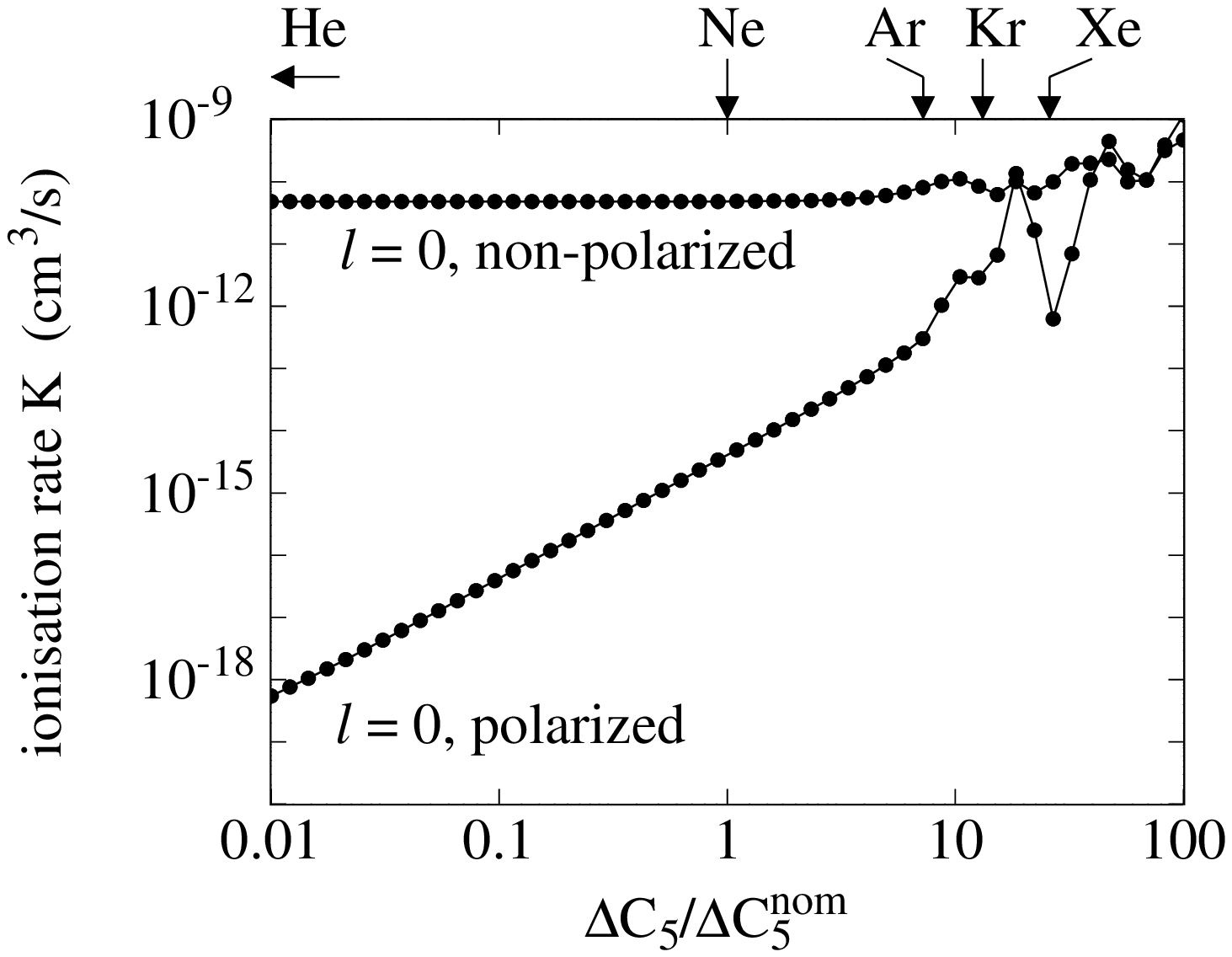,width=13cm}
\caption{\label{figKcsix}
Partial ionization rates for trapped \Netpt-atoms as a 
function of the normalized long-range anisotropy parameter
\deltacfive/\deltacfivenom. A quadratic dependence of the
rate for polarized atoms on the anisotropy parameter
is observed. Arrows indicating the estimated anisotropy
parameter for \meta He, Ne through Xe 
(listed in Table \ref{tabledD}),
illustrate the decrease in ionization suppression ratio
with increasing rare gas atom mass.
The collision energy is $E = 1 \mu\Kelvin$.
} 
\end{figure}

A convenient way of representing the variation in \cfive is by
introducing a parameter \deltacfive\  = $C_5^{max}$-$C_5^{min}$ equal to
the difference of the \cfive coefficients of the most and least
attractive long-range potentials. Table (\ref{tableMolec}) gives
\deltacfive\ = 0.84 a.u.\ for Ne, which may be considered its
``nominal'' value, which we will denote by \deltacfivenom. We can then
study the influence of long-range anisotropy by calculating the
ionization rates as a function of \deltacfive/\deltacfivenom. Such a
calculation is shown in Fig.\ (\ref{figKcsix}), where, for clarity, we
have omitted any anisotropy in the DD-term by using an identical,
averaged \csix-coefficient for all states involved. The figure points to
a quadratic dependence of $K^{\text{pol}}$ on the dominating anisotropy, while
$K^{\text{unpol}}$ is basically unaffected by the change in \cfive.
This result can be explained by simple qualitative arguments. The
anisotropic long-range interaction \cfive/$R^5$ will cause a rotation of
the molecule's electronic angular momentum $j$ over an angle that is
proportional to the variation \deltacfive\ in \cfive, as follows from
the equation of motion $I\mbox{d}\Omega/\mbox{d}t=\tau$ for the
classical rotation angle $\Omega$ of an object with a moment of inertia
$I$ acted upon by a torque $\tau$ (for an anisotropic potential, $\tau$
is proportional to $\mbox{d}V(R)/\mbox{d}\Omega$). In a binary collision
of two polarized atoms, by analogy the initially pure quintet state will
develop a projection $\mid P \rangle$ proportional to $\deltacfive$ on
other, ionizing states. Since the probability for ionization in such a
collision will be proportional to $\langle P \mid P \rangle$, we obtain
the quadratic dependence of $K^{\text{pol}}$ on $\deltacfive$ found in
Fig.\ (\ref{figKcsix}).

From this figure we can therefore conclude that the limit on the
suppression ratio set by long-range anisotropy quickly becomes stricter
as the rare gas becomes heavier, because \deltacfive\ is proportional to
the square of the quadrupole moment of the core (see Eq.\
(\ref{eqCsix})). We have listed the corresponding matrix elements in
Table \ref{tabledD} for He through Xe, based on values taken from Ref.
\cite{froese}, and indicated the relative anisotropies of all the rare
gases in Fig.~\ref{figKcsix}. For He, $\deltacfive = 0$ since only
$s$-electrons are involved so that no limit on \rs results from the
long-range anisotropy. (For He therefore, a first limit is set by the
spin-dipole interaction \cite{fedichev,fedichev1}). For Ar through Xe,
however, $\deltacfive$ is appreciably larger than for Ne, suggesting
that much smaller suppression ratios should be expected for these
heavier atoms. Indeed, the available experimental evidence for
metastable Kr~\cite{krexpts} and Xe~\cite{xeexpts}, 
with, according to Table \ref{tabledD}, approx.\ 200 and 700 times
larger $\deltacfive^2$ than Ne, respectively,  
showed basically no suppression at all.

\section{Comparison with metastable xenon experiments.}

The most extensive set of data is available for Xe,
where recent experiments \cite{xeexpts} suggest that spin-polarization
does not lead to suppression of the ionization rate for any of the
available isotopes; for all bosonic isotopes,
spin-polarization even seems to increase the ionization rate. From the
discussion in Subsection~\ref{secVarC} one would already expect \rs to
be less than 10 in this case. However, additional dynamical effects due
to the larger mass of Xe, and the difference in short-range potentials
compared to Ne should also be taken into account. These changes can be
easily incorporated into our model by substituting Cs potentials
\cite{servaas} for the Na potentials used for Ne, and using the
appropriate mass. We have used the same auto-ionization widths as for
Ne, however, since auto-ionization widths for Xe are not available.

\begin{figure}
\epsfig{file=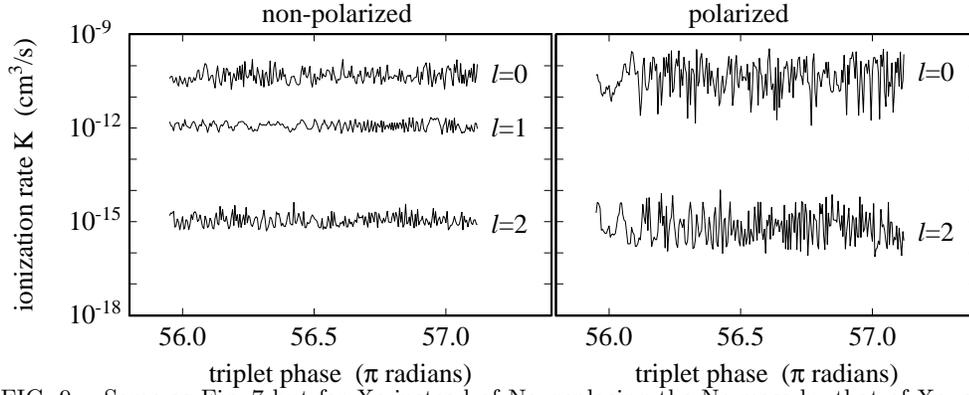,width=13cm}
\caption{\label{figKxe}
Same as Fig.~\protect\ref{figVarT} but for Xe instead of Ne, replacing
the Ne mass by that of Xe, and the Na-potentials by those of Cs.
The larger anisotropy in the case of Xe leads to much greater
polarized ionization rates than for Ne, often exceeding those
for unpolarized atoms. A suppression $\eta = 10$ is possible,
but values $\eta < 1$ are more likely. The collision energy is
$E = 1 \mu\Kelvin$.
} 
\end{figure}

Figure~\ref{figKxe} shows the result of such a treatment for a collisions
energy of $1 \mu$K. Again we varied the amount of phase available in the
triplet-potential to obtain the variation of the ionization rates for
polarized and unpolarized clouds of (Xe) atoms. Indeed, suppression by a
factor of $\rs \approx$10 is still possible, but over 60\% of the
$\phi$-range in Fig.~\ref{figKxe}, $\rs < 1$. In fact, the
$\phi$-averaged value is now $\langle \rs \rangle = 0.7$, showing that
for Xe ionization suppression by spin-polarization is very unlikely.
Here, the QQ-term mixes up the polarized and unpolarized states so
thorougly at long-range that they basically ionize at the same rate. The
fact that $\rs < 1$ can then be considered a quantum-statistical effect:
for polarized atoms, the $\ell = 0$ partial wave necessarily represents a
greater flux than for unpolarized atoms, since in the former case the
$\ell = 1$ partial wave is absent.

\section{Discussion.} \label{secConclusion}

In this paper we have developed a semi-quantitative description of
ionization in metastable neon collisions at very low temperatures.
In fact, this description holds for all the heavy metastable rare
gas atoms, but we have concentrated on neon because of ongoing
experiments in our laboratory as well as the availability of the
properties needed as input to the model.
We have put special emphasis on obtaining a lower limit on the ionization
rate for spin-polarized atoms, in order to investigate the possibilites
of reaching Bose-Einstein Condensation with \meta rare gas atoms.
While we are not able to provide a definitive answer, we do provide a
framework of potentials and auto-ionization widths that allow one
to investigate the system in a qualitative way. It has
become clear that such a limit depends strongly on the occurrence of
quasi-bound resonances in the potentials for fully polarized atoms with
$\ell > 0$. While it is possible to obtain reasonable estimates of the
anisotropy of the long-range part of these potentials, sufficiently
accurate information on the short-range parts is not available at this
moment. Such potentials may become available in the near future,
however \cite{svetlana}. Our calculations show
that, under favorable conditions, the long-range anisotropy still allows
for very effective suppression (up to $10^4$ times) of ionization by
spin-polarization, at least for \meta Ne. This suppression becomes less
effective as the atom becomes heavier, with a limit of $\rs \approx
10$ for Xe.

Additional couplings may exist between the quintet states and the
ionizing states that we have not considered. Such couplings may occur at
smaller $R$, where core-valence interactions become more important. The
interactions that will have an effect on the ionization rate for
spin-polarized atoms must create a direct coupling between a quintet
state and other, ionizing states. For Xe, e.g., one should expect a
large effect of the second-order spin-orbit splitting from the
experience with Cs~\cite{boesten,mies}. A simple splitting of the various
quintet potentials, however, due to a dependence of the potentials on a
$\sigma$ or $\pi$ orientation of the cores, does not have such an
effect, at least not for small magnetic confining fields. This is
because \meta rare gas atoms in general lack hyperfine-structure, so
that exothermic exit channels do not exist. As a result, depolarization
through such a mechanism is prohibited. We have checked this by adding a
phenomenological $\sigma/\pi$ core-splitting to our potentials, with an
order of magnitude comparable to that found in the Ne+Ne$^+$-system
\cite{neneplus}. The ionization rate for polarized atoms was found to
increase only slightly with the magnitude of this splitting.

A large magnetic confining field, however, could cause Zeeman splittings
greater than the centrifugal barrier present in the $\ell > 0$ quintet
states. This would open up new channels for depolarizing collisions that
require only a splitting within the quintet states. Given the height of
the centrifugal barrier, 5.8~mK for Ne and $\ell = 2$, such processes
should be expected to be fully developed at magnetic fields of
approximately 60~Gauss. It may be profitable, therefore, to use the
minimum practical trapping field for metastable rare gas atom traps.

\begin{figure} 
\epsfig{file=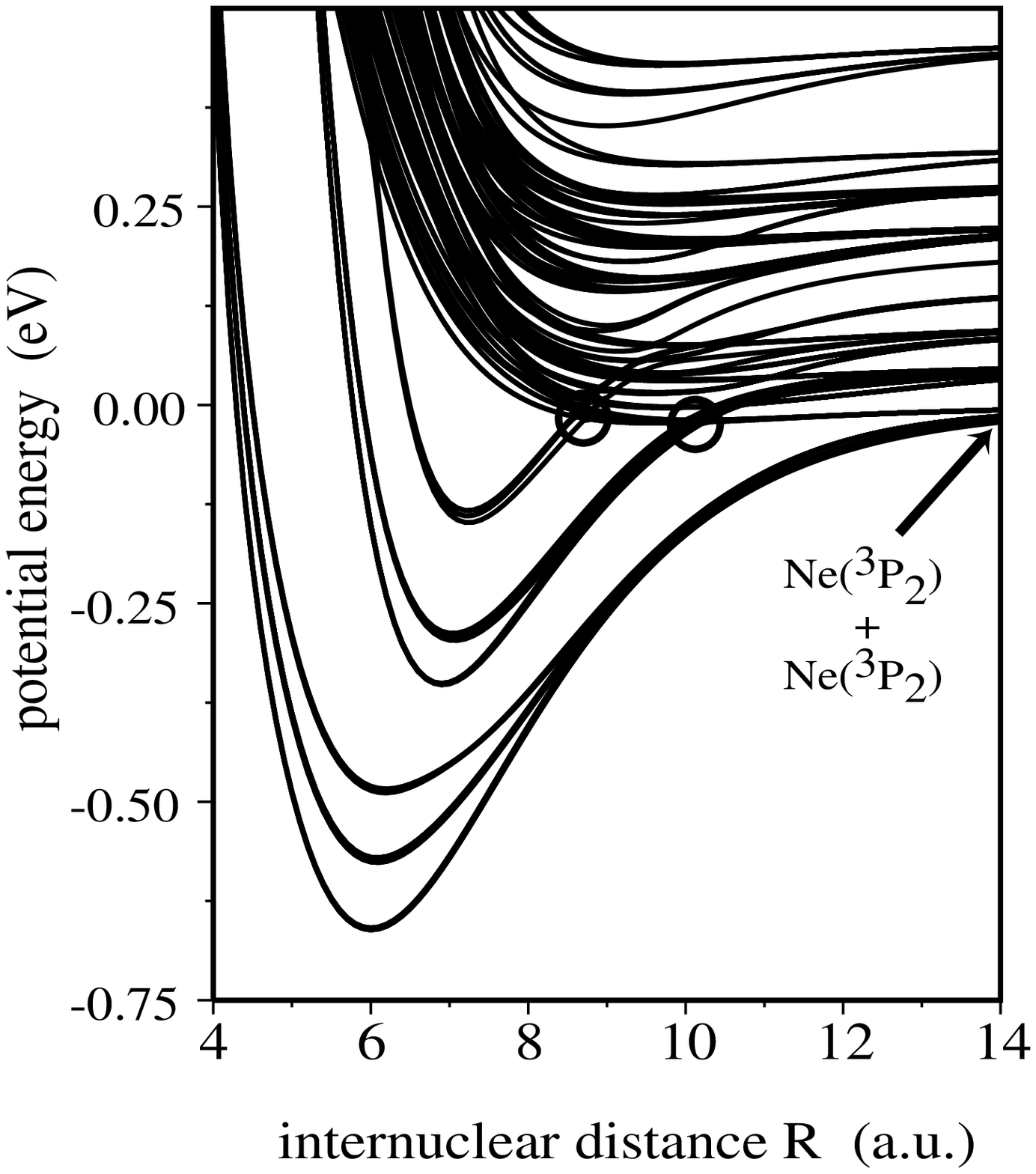,width=10cm}
\caption{\label{figFullpots} 
Adiabatic potentials for the Ne\configNe+Ne\configNe-system
including all fine-structure states. The zero-point of the energy
scale corresponds to the asymptotic energy of the
\Netpt+\Netpt-configuration. Compared to the
reduced set of Fig.\ (\ref{figPots}), deeper wells (resembling
more closely the \NaS\ state) have
developed in the states connected asymptotically to
\Netpt+\Netpt, indicating the relaxation of the asymptotic
spin-orbit configuration. Crossings of the
\Netpt+\Netpt-system with other asymptotic configurations
are indicated as well. Such crossings may lead to reduced
ionization suppression by funneling flux onto ionizing 
potentials.
} 
\end{figure}

Another factor that we have not taken into account is that
curve-crossings of the \Netpt+\Netpt\ states can occur with other
molecular states from the Ne\config+Ne\config\ system at some
intermediate range. For Ne, it is possible to develop potentials for
this full system with the recipe developed in Sec.\ \ref{secPots}. The
result of such a treatment is shown in Fig.\ (\ref{figFullpots}). In
this case, the \Netpt+\Netpt\ quintet states indeed cross several states
with other asymptotic limits. Closer inspection has taught us that some
of these are indeed avoided crossings, with the coupling due to the DD-
and QQ-interactions. At the crossing points, a second order treatment of
the DD-interaction in fact no longer suffices. Flux entering on one of
the quintet states can now be diverted onto curves that allow ionization
to occur at short range. In order for the effect of such crossings to be
taken into account, an $R$-dependent treatment of the DD- and
QQ-interactions would be necessary, and in addition a much larger system
of coupled equations would have to be solved. We consequently leave such
a treatment to a future paper.

\section{Acknowledgements.}

We gratefully acknowledge the support of a Cray Research Grant from Cray
Research, Inc., and the Dutch Organization for Fundamental Research
(NWO)/Dutch National Supercomputing Facilities (NCF). The work of
E.J.D.\ Vredenbregt is made possible by a fellowship from the Royal
Netherlands Academy of Arts and Sciences (KNAW). Additional support from
the Netherlands Foundation for Fundamental Research on Matter (FOM) is
also acknowledged. We thank J.G.C.\ Tempelaars, J.P.J.\ Driessen, T.H.
Bergeman, E.\ Tiesinga and C.\ Orzel for helpful discussions.

\newpage

\bibliographystyle{prsty}

\begin{thebibliography}{10}
\bibitem{neonprops}
A large set of references to papers dealing with the properties of neon is
  given in Ref.\ \protect\cite{ourpra}. See also Refs.\
  \protect\cite{c5ref,ddvalues}.
\end{thebibliography}

\newpage

\begin{table} \caption{\label{tableStates}
Spectroscopic notation, equivalent Russel-Saunders $LSJ$ notation where
applicable, atomic energy and electronic angular
momentum quantum number $J$ for the four states of
Ne in the \configNe-configuration
}
\begin{tabular}{ccdc}
Spectroscopic & $LSJ$ & Energy & $J$ \\
notation & notation & (cm$^{-1}$) & \\
\tableline
3s$^\prime$ [1/2]$_1$ & - & 135888.7137 & 1\\
3s$^\prime$ [1/2]$_0$ & $^3$P$_0$& 134818.6405 & 0\\
3s [3/2]$_1$  & - & 134459.2871 &1\\
3s [3/2]$_2$  & $^3$P$_2$ & 134041.8400 & 2\\
\tableline
\end{tabular}
\end{table}

\begin{table} \caption{\label{tableMolec}
Hund's case ($c$) symmetries, \csix and \cfive coefficients
of the 25 adiabatic molecular states asymptotically connected to \Netpt+\Netpt,
derived from the \NaS and \NaT-potentials with the formalism of
Sec.\ \ref{secPots}. The degeneracy of all $\Omega = 0$ states is 1,
that of all others is 2.
}
\begin{tabular}{ldd}
$\Omega$ & $C_5$ (au) & $C_6$ (au)\\
\tableline \\
3 & -0.3700 & -1953.24 \\
2 & -0.2878 & -1951.33 \\
2 & -0.2673 & -1952.60 \\
1 & -0.2028 & -1951.23 \\
0$^-$ & -0.1525 & -1951.64 \\
1 & -0.1318 & -1951.13 \\
0$^+$ & -0.0679 & -1954.13 \\
3 & 0.1233 & -1953.36 \\
2 & 0.1233 & -1952.73 \\
1 & 0.1617 & -1952.78 \\
1 & 0.1730 & -1953.52 \\
0$^+$ & 0.2173 & -1955.58 \\
4 & 0.2467 & -1955.54 \\
0$^-$ & 0.3992 & -1951.48 \\
0$^+$ & 0.4673 & -1952.38 \\
\end{tabular}
\end{table}

\begin{table} \caption{\label{tabledD}
Atomic and core polarizability, \dval\ and \dcore, respectively, and
expectation value of the square of the orbital radius of the hole-state
in the \pcore-core, (taken from
Ref.~\protect\cite{froese}) for \meta rare gas atoms Ne through Xe. The
atomic polarizability corresponds to the value listed by Ref.\
\protect\cite{ddvalues} for configuration \config, while the core
polarizability corresponds to the ionic configuration \pcore. Since
$\dcore/\dval \ll 1$ always, one may equate the polarizability of the
$(n+1)$s-valence electron alone with the atomic polarizability. The last
two columns give the anisotropy parameters \deltacsix\ and
\deltacfive\ relative to those for
\Netpt ($\deltacfive = 0.84$ a.u., $\deltacsix = 4.5$ a.u.).
For clarity, the table also lists the anisotropy parameters for
He as being zero: in this case the core has a spatially isotropic (1s)
configuration.
}
\begin{tabular}{ccddddd}
rare gas & principal quantum & \dval & \dcore &  $\langle r^2_h \rangle$
& anisotropy \deltacsix & anisotropy \deltacfive \\
atom & number $n$ & (a.u.) & (a.u.) & (a.u.) & relative to \Netpt &
relative to \Netpt\\
\tableline
He &   & & & & 0 & 0 \\
Ne & 2 & 153 & 1.33 & 1.23 & 1.0 & 1.0 \\
Ar & 3 & 305 & 7.90 & 3.31 & 3.0 & 7.3 \\
Kr & 4 & 363 & 12.8 & 4.46 & 4.1 & 13.2 \\
Xe & 5 & 511 & 23.1 & 6.28 & 5.2 & 26.1 \\
\end{tabular}
\end{table}

\newpage

\newpage

\newpage

\newpage

\newpage

\end{document}